\documentclass{aa}
\usepackage{txfonts}
\usepackage{graphicx}

\newcommand{\s}{\scriptsize}

\begin{document}
\title{Subhaloes in $\Lambda$CDM cosmological simulations}

\subtitle{I. Masses and abundances}

\author{P. Nurmi\inst{1}\fnmsep\thanks{pasnurmi@utu.fi}, P.
Hein\"am\"aki\inst{1},
E. Saar\inst{2}, M. Einasto\inst{2}, J. Holopainen\inst{1},
V. J. Mart{\'\i}nez\inst{3}  and J.
Einasto\inst{2}}

\offprints{P. Nurmi}

\institute{Tuorla Observatory, University of Turku, V\"ais\"al\"antie
20, FI-21500 Piikki\"o,
Finland
\and
Tartu Observatory, T\~oravere, Tartumaa, 61602 Estonia
\and
Observatori Astron\`omic, Universitat de Val\`encia, Apartat de Correus
22085, E-46071 Val\`encia, Spain }
\date{}

   \abstract
    {}
    {If the concordance $\Lambda$CDM model is a true description
      of the universe, it should also properly predict the
      properties and structure of dark matter haloes, where
      galaxies are born.
    Using N-body simulations with a broad
   scale of mass and spatial resolution, we study the structure
   of dark matter haloes, the distribution of masses and the
    spatial distribution of subhaloes within the
   main haloes.}
    {We carry out three $\Lambda$CDM simulations with different
       resolutions using the AMIGA code.
       Dark matter haloes are identified using an algorithm
       that is based on the adaptive grid
       structure of the simulation code. The haloes we find encompass
    the mass scales from $10^8\mathrm{M}_{\sun}$ to
$10^{15}\mathrm{M}_{\sun}$.}
    {We find that if we have to study the halo structure (search for
    subhaloes), the haloes have to contain at least $10^4$
    particles. For such haloes, where we can resolve substructure,
    we determined the subhalo mass function and found that it
    is close to a power law with the slope $-0.9$ (at present time),
    consistent with previous studies. This slope depends slightly on
    the redshift and it is approximately the same for main haloes.
    The subhalo mass fraction ($M_{\mbox{\s subH}}/M_{\mbox{\s MH}}$) 
    is between 0.08 and
    0.2, increasing slightly with redshift and with
    the mass of the main halo. Its distribution is approximated using 
the Weibull distribution at different epochs. 
The mean values of subhalo mass are independent of the main halo
mass. The spatial density of subhaloes, scaled to the virial radius
of the main halo ($r_{\mbox{\s vir}}$), is independent of redshift and follows the
$r^{1/3}$ rule. }
{}


\authorrunning{P. Nurmi et al.}
\titlerunning{Subhaloes in $\Lambda$CDM cosmological simulations}

\maketitle

\section{Introduction}

A recent remarkable achievement in cosmology is due to the NASA's
Wilkinson Microwave Anisotropy Probe (WMAP) measurements of the CMB
fluctuations (Bennet et al. \cite{bennet}, Spergel et al.
\cite{sperge}). The agreement of the theoretical predictions with
the angular power spectrum of fluctuations measured by WMAP,
together with the results of other diverse cosmological studies
(models of the nucleosynthesis and light element abundances,
supernovae data, and large-scale structure observations etc.), seem
to favor a simple $\Lambda$CDM concordance model. This gives us a
well-defined and pretty well restricted (in sense of free
parameters) base for studies of formation of the observed structure
(galaxies, stars, etc.).

Despite the great success of the  concordance model there are
several open questions, ranging from fundamental ones, as the nature
of dark matter and dark energy, to more specific problems related to
inflationary models and structure formation. In this paper we will
concentrate at one specific structure formation problem, the
properties of dark matter haloes and their subhaloes.

The best-known observational counterparts of massive dark haloes are
clusters of galaxies. These are the most massive and the largest
gravitationally bound systems known to exist in the Universe. Being
the vanguard in non-linear regime, clusters of galaxies are
important link between the initial density field and present day
structures in the Universe. Resent numerical and analytical studies
of the cluster scale dark matter (DM) haloes agree well with
observed cluster abundances (Press \& Schechter \cite{press},
Jenkins et al. \cite{jen:jen} and Sheth \& Tormen \cite{sheth}).

As a subsequent step, high-resolution numerical studies are pushing
the theory of the structure formation to smaller scales toward the
'galactic' subhalo region (subhaloes are haloes within the virial
radius of the main halo) (De Lucia et al. \cite{lucia}, Diemand et
al. \cite{diemand}, Gao et al. \cite{gao}, Gill et al.
\cite{gill-I}).

This is an important step because the substructure of large DM
haloes links cluster haloes and galaxy haloes together, providing
observable probes for structure formation scenarios. Probably the
most difficult problem at small scales is the so-called 'dwarf
galaxy crisis'; simulations predict substantially more substructure
(about two orders of magnitude) within the galactic DM haloes than
observed (Moore et al. \cite{moore2} and references therein).

Several studies have shown that halo substructure can substantially
affect the observed flux ratios of gravitationally lensed quasars
(Bradac et al. \cite{bradac} , Metcalf \& Madau \cite{metcalf}, Chen
et al. \cite{chen}). Mao et al. (\cite {mao2}) conclude that
anomalous flux ratios in lenses require that the surface mass
density fraction in substructures at typical image positions is a
few percent. This is higher than the surface density value predicted
by the $\Lambda$CDM model (about half a per cent). The required
substructure masses are $10^4$--$10^8$~M$_{\sun}$. This is an
obvious challenge for numerical simulations. For weak-lensing
studies the problem may be complicated yet by the badly known mass
profile (this is often approximated by the simple NFW profile,
extrapolated to distances well beyond $r_{\mbox{\s vir}}$). However,
according to Prada et al. (\cite{prada}) the density profiles of the
dark matter haloes beyond the formal virial radius differ
considerably from the NFW profile.

In order to encompass a large enough volume, and to obtain
sufficient mass resolution, we carried out three $\Lambda$CDM
cosmological simulations for different mass resolutions and volumes.
Comparing the simulations, we can estimate the resolution effects,
and can find resolution-independent properties of substructure.
Specifically, we study the subhalo content of haloes and find the
mass and number distributions of subhaloes. We also study the
distributions of mass fractions and discuss evolution of
substructure, and analyze the spatial distribution of subhaloes in
an around their main haloes.

\section{Simulations}
For the present study we use a flat
($\Omega_m+\Omega_{\Lambda}+\Omega_b=1$) cosmological background
model with the parameters derived by the WMAP microwave background
anisotropy experiment team (Bennett et al. \cite{bennet}): the dark
matter density $\Omega_m=0.226$, the baryonic density
$\Omega_b=0.044$, the vacuum energy density (cosmological constant)
$\Omega_{\Lambda}=0.73$, the Hubble constant $h=0.71$ (here and
throughout this paper $h$ is the present-day Hubble constant in
units of 100 km s$^{-1}$ Mpc$^{-1}$) and the rms mass density
fluctuation parameter $\sigma_8=0.84$. The transfer function and the
initial data for our models were computed using the COSMIC code by
E.~Bertschinger (\texttt{http://arcturus.mit.edu/cosmics/}).

Each N-body integration algorithm has its advantages and weaknesses,
so arguably none of them is completely satisfactory. Increasing
resolution of the simulations increases also the requirements for
the N-body code and for the analysis tools. Thus to avoid
computational artifacts in the results, it is essential that
different codes are used. This makes it possible to cross-check the
results of the complex dynamics of the cosmic structure formation.
Moreover, additional realizations are always needed, to improve the
'N-body statistic' that is pretty poor due to heavy CPU and memory
requirements of the simulations of substructure scales.

The simulations presented here were carried out using the AMIGA code
(Adaptive Mesh Investigations of Galaxy Assembly) that is the
updated version of the MLAPM code by Knebe et al. (\cite{knebe}),
(\texttt{http://www.aip.de/People/aknebe/AMIGA/}). The AMIGA code is
adaptive, with subgrids being adaptively formed in regions where the
density exceeds a specified threshold.

For the halo identification we adopt a relatively new method from
the AMIGA toolbox, called MHF (Gill et al. \cite {gill-I}), that is
based on the adaptive grid structure of the AMIGA. The centroids of
the densest grid volumes (at the bottom of the grid tree) are used
as the halo centers. From this point, radial bins are followed
outwards until the radius $r_{\mbox{\s vir}}$ where the density level
$\rho_{\mbox{\s satellite}}(r_{\mbox{\s vir}})=\Delta_{\mbox{\s vir}}(z)\rho_{b}(z)$,
where $\rho_b$ is the mean cosmological density and
$\Delta_{\mbox{\s vir}}(z)$ is the overdensity for virialized objects.
Another possibility to evaluate the size of the halo is to find the
distance $r_{\mbox{\s trunc}}$, where the radial density profile starts
to rise. Hence, the outer radius of the halo is either $r_{vir}$ or
$r_{trunc}$, whichever is smaller. The properties of the halo (mass,
shape etc.) are calculated for all bound particles inside this
limit.

According to the MHF procedure, subhaloes are virialized objects
inside the  virial radius of main haloes. By using the refinement
hierarchy to trace gravitationally bound objects,  MHF gives an
efficient way to extract haloes-within-haloes. This is not so easy,
for example, for the widely used FoF-method. Gill et al.
(\cite{gill-I}) compared halo identification by MHF against other
two popular methods, SKID and FoF. With a suitably chosen linking
length, MHF and SKID give very close results, but MHF is not as
sensitive in finding subhaloes in the central regions of main haloes
($r/r_{\mbox{\s vir}}<0.2$) as SKID or FoF. The best results could be
obtained by tracking satellites, but this is out of the scope of
this paper. The most important practical difference between MHF and
other halo finders is that one does not have to assume a linking
length in MHF. In principle, MHF could also find sub-sub structure,
but our mass resolution is not sufficiently good for such a detailed
analysis. The MLAPM and MHF codes have been used previously  for
subhalo studies, by Gill et al. (\cite {gill-I}), Gill et al. (\cite
{gill-II}) and Gill et al. (\cite {gill-III}).

For the present study we carried out three different simulations
(designated B10, B40, B80, according to the box size) with different
box sizes and resolutions. This is a compromise between our computer
resources and the mass resolution; in this way we obtain dark matter
haloes at wide range of masses: from $10^{15} \mathrm{M}_{\sun}/h$
down to $10^8 \mathrm{M}_{\sun}/h$. The parameters of the
simulations are summarized in table~\ref{tab1}, where $L$ is the
size of the box and $z_i$ is the starting redshift of the
simulation.

The smallest simulation box is only 10 Mpc/h big and one may argue
that the large scale modes, ignored in this very small volume
simulation, can cause spurious errors in the results. Recently,
Bagla \& Prasad (\cite{bagla}) analyzed the effects of finite
simulation box on halo mass functions. They found that the main
effect is that the abundance of high mass haloes is underestimated
and the number of haloes of smaller mass might be overestimated. In
general, the errors are small, if the scales of interest are
sufficiently smaller than the box size. A similar conclusion was
obtained by Power \& Knebe (\cite{power}), who showed that the
number of intermediate mass haloes
($M\sim10^{13}\mathrm{M}_{\sun}/h$) is overestimated, but the
high-mass haloes are suppressed, if the long wavelength
perturbations are neglected. However, the distributions of
concentrations remain the same. Hence, if we restrict our analysis
to intermediate mass haloes and their subhaloes, then also the
smallest B10 simulation gives reliable results. If we look carefully
at the B10 mass function (Fig.~\ref{difmf1}), we can notice a slight
overabundance of intermediate haloes (around
$M\sim10^{12}\mathrm{M}_{\sun}/h$) with respect to the B40 and B80
simulations.

\begin{table*}
 \caption{Summary of the simulation parameters}
 \label{tab1}
 \centering
  \begin{tabular}{cccccc} \hline \hline
  Simulation & L [Mpc/h]& Number of particles & Mass resolution
  [$\mathrm{M_{\sun}/h}$] & Force resolution [kpc/h] & $z_i$ \\
\hline
B10 & 10 & $256^3$ & $4.47\times10^6$ & 0.46 & 71.52\\
B40 & 40 & $256^3$ & $2.86\times10^8$ & 1.8 & 47.96\\
B80 & 80 & $256^3$ & $2.29\times10^{9}$ & 7.3 & 38.77\\
\hline
\end{tabular}
\end{table*}

\begin{figure}
 \centering \resizebox{\hsize}{!}{\includegraphics*{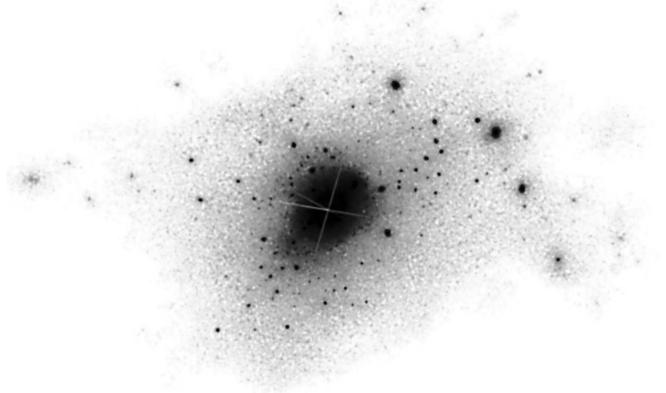}}
 \caption{A typical large
 $M=10^{12}\mathrm{M}_{\sun}/h$ main halo with surrounding
subhaloes. } \label{fig:subhaloes}
\end{figure}

Our haloes are bound structures identified by the MHF halo finder
algorithm. We divide them into four categories:
\begin{itemize}
\item main haloes (haloes with subhaloes) (MH);
\item single haloes (haloes without identified subhaloes) (SH);
\item subhaloes (bound structures inside the virial radius of the
main halo) (SubH);
\item all previous halo classes together (All).
\end{itemize}

Table ~\ref{tab2} shows the total numbers of haloes according to the
classification above, and the percentages of each halo type. Two
points should be noted, both of which will be discussed in later
sections. The fraction of main haloes with respect to all haloes is
nearly constant in all three simulations. The fraction of subhaloes
increases rapidly with resolution, indicating how a better
resolution reveals a more detailed structure. The majority of all
haloes are single, but this fraction becomes smaller as the
resolution increases.

\begin{figure}
 \centering \resizebox{\hsize}{!}{\includegraphics*[angle=-90]{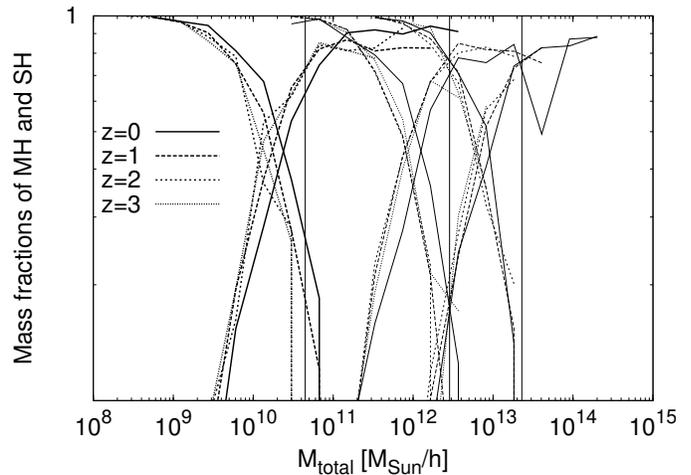}}
 \caption{Mass fractions of
main haloes and single haloes with respect to the total mass, for
different total halo mass intervals. Data for four redshifts ($z=0,
z=1, z=2, z=3$) are shown. The descending curves are for single
haloes and the ascending curves are for main haloes. The three
groups of curves refer to the three simulations (B10, B40, B80; from
left to right) and the three vertical lines show the resolution
limits for main haloes.} \label{fractions_MH_SH}
\end{figure}

Only a fraction of all mass particles forms bound structures
(haloes), and single haloes represent the majority of haloes for all
resolutions. The existence of single haloes is certainly a
resolution effect, but it is still unclear if all haloes would
actually harbor subhaloes, if the resolution of the simulations
would be ideal. The behavior of the mass fractions of main haloes
and single haloes with respect to the total mass in all haloes
(Fig.~\ref{fractions_MH_SH}) clarifies the single haloes problem.
The three ascending groups of curves show the mass fractions of main
haloes and the descending groups of curves show the mass fractions
of single haloes. The vertical lines show the (lower) resolution
limits, explained in the next section, of the simulations. We see
that in all three simulations the mass fraction of single haloes
decreases rapidly, as the resolution limit is reached. Beyond this
limit there are hardly no single haloes at all, and the main halo
mass fraction is constant (0.8-0.9) in this region (the remaining
haloes are subhaloes). Thus it is difficult to estimate the real
number of single haloes, which are too small to have substructure;
probably there are none.

\begin{table*}
 \caption{Number of haloes of different types}
 \label{tab2}
 \centering
 \begin{tabular}{cccccccc}
 \hline \hline\\[-8pt]
 Simulation & Main haloes & Subhaloes & Single haloes & All haloes &
 $\frac{\mbox{Main haloes}}{\mbox{All haloes}}[\%]$ &
 $\frac{\mbox{Subhaloes}}{\mbox{All haloes}}[\%]$ &
 $\frac{\mbox{Single haloes}}{\mbox{All haloes}}[\%]$  \\[6pt]
 \hline
B10 & 168 & 829 & 3854 & 4851&3.5&17&79\\
B40 & 287 & 892 & 5657 & 6836&4.2&13&82\\
B80 & 332 & 629 & 8398 & 9359&3.5&6.7&89 \\
\hline
\end{tabular}
\end{table*}

\section{Properties of the haloes}

\subsection{Mass functions}

As the halo mass function can be predicted theoretically (Press \&
Shechter \cite{press}), Sheth \& Tormen (S-T, \cite{sheth}), it
provides an important observational constraint on the parameters of
the cosmological model and on the amplitude of initial fluctuations.
The theoretical predictions have been checked by N-body models by
many authors and have been found to work well (see, e.g., Gao et al.
\cite{gao} and references therein). From the observational side,
masses of galaxy clusters can be derived using either X-ray data and
the mass-temperature relations, or data from optical surveys, using
the velocity dispersion of galaxies in clusters (virial masses).
Since obtaining the cluster masses empirically is not an easy task,
only a few observational cluster mass functions have been found
(Bahcall and Cen (\cite{bac}), Biviano et al. (\cite{biv}), Reiprich
\& B\"ohringer (\cite{rei:rei}), Girardi and Giuricin
(\cite{gir:giu}) and Hein\"am\"aki et al. (\cite{hei:hei})).

\begin{figure}
 \centering
 \resizebox{\hsize}{!}{\includegraphics*[angle=-90]{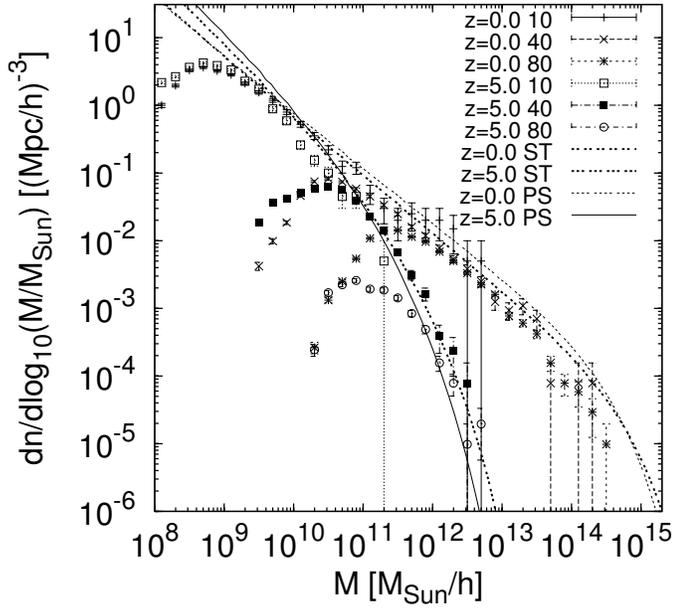}}
 \caption{Differential mass functions of all
haloes in three simulations at two different redshifts $z=0$ and
$z=5$ (see the legend in the Figure). The theoretical
Press-Schechter (PS) and Sheth \& Tormen (ST) predictions are also
shown.} \label{difmf1}
\end{figure}

Usually, the mass function of galaxy clusters/groups is defined as
the number density of clusters above a given mass $M$,
\mbox{$n(>M)$}. This is useful if we are mainly interested in the
total number density of clusters. We represent the mass function in
this paper by its differential form, $dn/dM$, that shows better
the behavior of the mass distribution at different scales. In
figure ~\ref{difmf1}, we plot the combined differential mass
functions of all haloes in three simulations B10, B40 and B80, for
two epochs, $z=0$ and $z=5$. We have also calculated the theoretical
predictions by the Press-Schechter (P-S) theory (Press \& Shechter
\cite{press}) and the Sheth \& Tormen (S-T, \cite{sheth}) theories,
using the same power spectrum $P(k)$ that was used for the initial
setup of the simulations. The Poisson error bars are also shown. We
see that the mass functions agree relatively well with the
analytical predictions for the halo abundances at different
redshifts. We have also compared our results with those of Gao et
al. (\cite{gao}) and Reed et al. (\cite{Reed2003}), and found that
the mass functions agree very well.

\begin{figure}
 \centering \resizebox{\hsize}{!}{\includegraphics*[angle=-90]{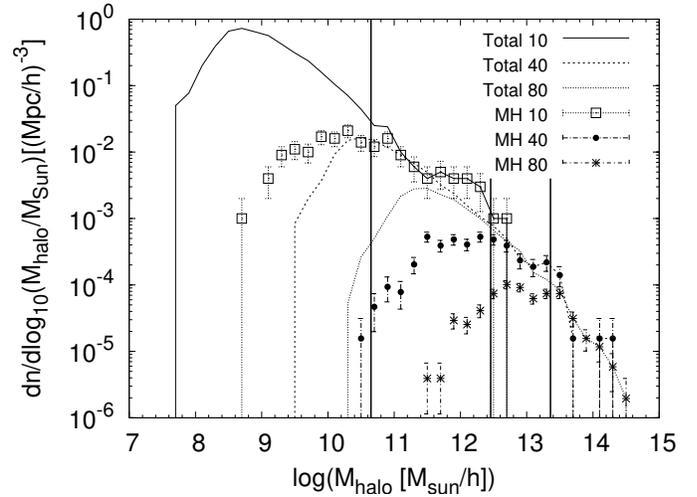}}
 \caption{Differential
mass functions of main haloes in three simulations are shown with
points. The errorbars are $1\sigma$ Poissonian. The mass functions
for all haloes are also given for comparison (lines). The vertical
lines mark the reliability limits, and only the main haloes that are
in the reliable region (the region to the right from the line) are
included in the analysis.} \label{Kuva3a_mf_MH_all}
\end{figure}

The differential mass functions of main haloes are shown in
Fig.~\ref{Kuva3a_mf_MH_all}, together with the mass functions for
all haloes. The main halo mass functions deviate clearly from the
general mass functions, after a certain halo mass value. This mass
gives the resolution limit of main haloes (about $N_r=10^4$
particles in all simulations). The corresponding masses for $N_r$
are $4.47\times10^{10}$ $\mathrm{M}_{\sun}/h$, $2.86\times10^{12}$
$\mathrm{M}_{\sun}/h$, and $2.29\times10^{13}$ $\mathrm{M}_{\sun}/h$
in the 10, 40 and 80 Mpc/$h$ simulations, respectively. Below these
mass limits the substructure is smeared out by numerical effects,
and the subhalo properties are not reliable. In the later analysis
we restrict ourselves only to main haloes with masses above these
resolution limits. For curiosity, we may plot all the data, but the
resolution limits are always shown as vertical lines.

\subsection{The subhalo mass function}

The mass function of subhaloes has been extensively studied lately.
Moore et al. (\cite{moore2}) suggested that the mass function of
substructures is independent of the mass of the parent halo.
Generally, the accuracy of the simulations or the small number of
subhaloes found have not permitted verification of this suggestion,
thus so far this is an open question. The subhalo mass function is
usually given in a simple exponential form:
\begin{equation}
\label{sublaw}
    dn/dm \propto m^{-\alpha}\quad
    \mathrm{  or  }\quad
    dn/dlog(m) \propto m^{\beta},\quad \beta=1-\alpha
\end{equation}
(for a restricted subhalo mass interval), where $\alpha$ does not
depend on the parent halo mass. Several studies have derived almost
the same slope values. De Lucia et al. (\cite{lucia}) estimated
that $\beta$ between $-0.94$ and $-0.84$ gives a good fit to their
data. This is very close to the value $-0.8$ in Helmi, White \&
Springel (\cite{helmi:helmi}) for a high resolution single-cluster
simulation. Also, Gao et al. (\cite{gao}) and Ghigna et al.
(\cite{ghigna}) obtained very similar values of $\alpha$, between
1.7 and 1.9. Hence, all the studies using different simulation
algorithms, mass scales and subhalo identification algorithms seem
to agree that the subhalo mass function can be well described by a
power-law of a single slope value, and the mass function does not
depend on the properties of the main halo (its mass). The subhalo
mass function is universal, depending only on the background
cosmology; also, the mass function is the same for
subhaloes and main haloes. However, there may still be a weak mass
dependence as suggested by Gao et al. (\cite{gao}) and Reed et al.
(\cite{reed}).

\begin{figure}
 \centering
 \resizebox{\hsize}{!}{\includegraphics*[angle=-90]{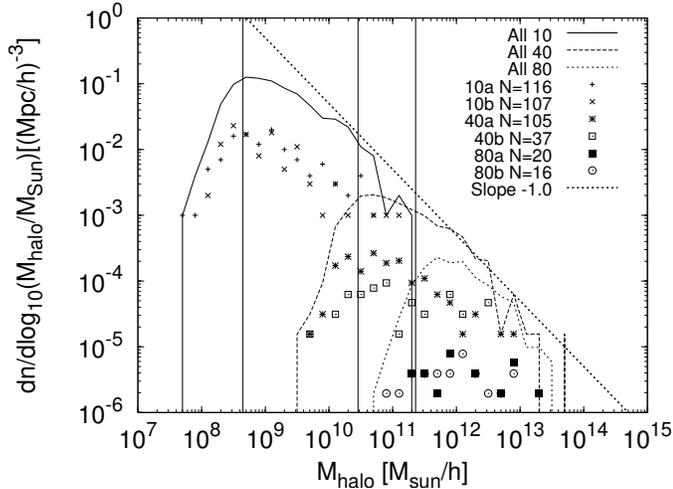}}
 \caption{Differential subhalo mass
functions for the three simulations.  The vertical lines show the
resolution limits of subhaloes (100 particles) in each simulation.
The points show the mass distributions for subhaloes surrounding the
two most massive main haloes. The straight line is the best fit for
all the haloes in the three simulations.} \label{Kuva4_shmf}
\end{figure}

We show the differential subhalo mass functions in the three
simulations Fig.~\ref{Kuva4_shmf}, together with the best fit
slope ($\beta = -0.9$). The vertical lines show the reliable regions
for subhaloes, where the subhaloes have at least ($N_h = 100$)
particles. This limit is not exact, but it is close to the point
after which the mass functions start to turn downwards, reflecting
incompleteness of the data.

In order to see if the differential subhalo mass functions of
individual main haloes differ from that of the total distribution,
we collected the two main haloes with the richest substructure from
every simulation and calculated the mass functions of the subhalo
populations. These are shown as points in Fig.~\ref{Kuva4_shmf}.
There are enough subhaloes to estimate these distributions only in
the B10 and B40 simulations. The numbers of subhaloes for the two
most abundant main haloes in these simulations are 116 and 107, and
105 and 37, respectively. Fig.~\ref{Kuva4_shmf} shows that the
individual mass functions also follow the general slope, indicating
that the differential subhalo mass distribution is universal.

\begin{figure}
 \centering
 \resizebox{\hsize}{!}{\includegraphics*[angle=-90]{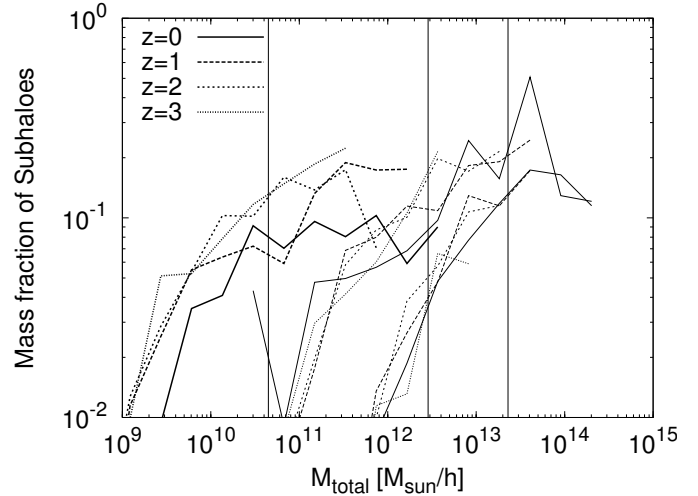}}
 \caption{Mass fractions of subhaloes
with respect to the total halo mass divided in different
mass intervals at four redshifts.
Three groups of curves refer to simulations B10, B40 and B80,
from left to right, respectively.}
\label{fractions_SubH}
\end{figure}

The halo mass and redshift dependence of the subhalo mass fraction
(with respect to the total halo mass) is shown in
Fig.~\ref{fractions_SubH}. The reliable mass regions for main haloes
can be seen in the figure as those where this mass fraction
practically does not depend on the total halo mass.

The mass fraction of subhaloes in the reliable region is between
0.08 and 0.2, it depends slightly on the total halo mass, and it
might depend on the redshift (see the curves for the model B10).
Since there are practically no single haloes in this region, the
mass fraction shown is the same as $\Sigma M_{SubH}/\Sigma M_{MH}$.

\subsection{Evolution of the subhalo mass function}

The theoretical halo mass function changes in time, as can be seen
in Fig.~\ref{difmf1}. To study this change more accurately, we fitted
power laws to the mass distributions (as in Fig.~\ref{Kuva4_shmf}).
Only haloes in the reliable regions were used in the analysis.

\begin{figure}
 \centering
 \resizebox{\hsize}{!}{\includegraphics[angle=-90]{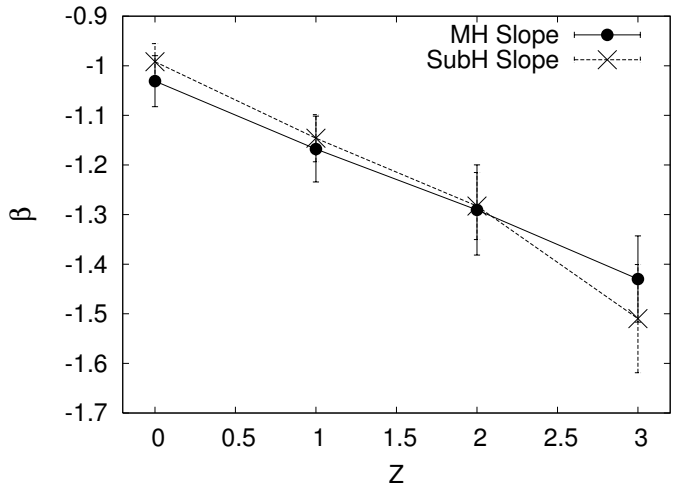}}
 \caption{Logarithmic slope
value ($\beta$) for the subhalo and main halo mass functions as a
function of redshift.} \label{z-evolution_mf_subh}
\end{figure}

The results of the (least square) fits are shown in
Fig.~~\ref{z-evolution_mf_subh}. The data from all the tree
simulations are used; the slope shown is $\beta$ in (\ref{sublaw}).

There is a reliable change in the slope, both for subhaloes and for
main haloes, from $-1.5$ at the redshift $z=3$ to $-1$ at $z=0$. Our
results agree with earlier studies for the redshift $z=0$ (e.g.,
DeLucia et al. (\cite{lucia})). However, we have found a notable
change in the slope value. The change is the same for main haloes
and subhaloes (within the errorbars), but the Figure hints that the
change in the slope of the subhalo mass function might be steeper.
The overall redshift dependence is linear; $\beta=-(1+0.15z)$ gives
a good fit to the data.

\subsection{Numbers and masses of subhaloes}

\begin{figure}
 \centering
 \resizebox{\hsize}{!}{\includegraphics*[angle=-90]{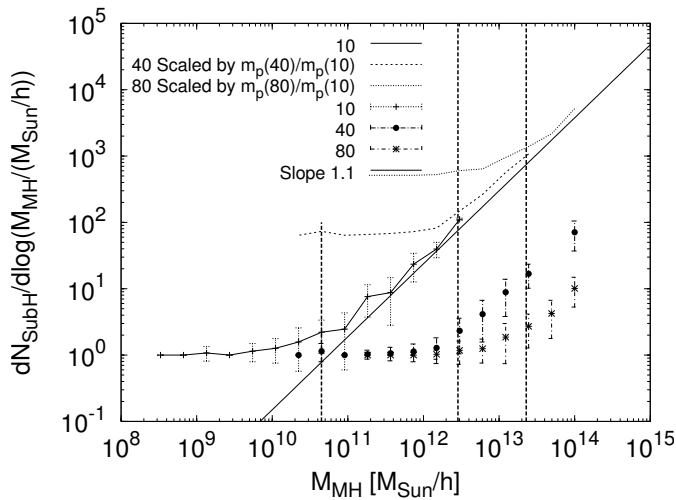}}
 \caption{Number of subhaloes as a
function of the main halo mass. The points show the raw simulation
results, with the error bars showing the standard deviations for
every bin. The lines show the scaled subhalo numbers. The straight
line shows the best fit relation, fitted to the points for the B10
simulation and to the scaled values for the B40 and B80
simulations.} \label{Kuva5_nofhaloes}
\end{figure}

Next we consider the mean number of subhaloes $N_h$ as a function of
the host main halo mass. This dependence is shown in
Fig.~\ref{Kuva5_nofhaloes} (for $z=0$). As the main halo mass
increases, its virial radius grows and it can host more subhaloes.
Because of the approximate self-similarity of the simulations with
different box sizes, the overall dependence can be calculated by
scaling the number of haloes by the ratio of the N-body particle
masses in different simulations. We found the best fit for the
subhalo number distribution as $N_h \propto M_{MH}^{1.1}$. This is
close to the linear relation found by Kravtsov et al. (\cite{kratso2})
in his halo occupation distribution (HOD) analysis. 
We calculated the
subhalo number distributions also for the earlier redshifts, up to
$z=3$, and found that the distribution is the same (within error
bars) as at $z=0$.

In every mass interval there is some scatter in the number of haloes
that is caused by the cosmic error. Fig. ~\ref{Kuva5_nofhaloes}
shows the standard deviations of the number of subhaloes for a given
main halo mass with error bars. The deviations are quite large for
some mass intervals, and for Milky Way size dark matter haloes ($\sim
10^{12} \mathrm{M}_{\sun}/h$) the mean subhalo number is 40 haloes
and the scatter $\sigma=10$. Our B10 simulation box is suitable to
study subhaloes around galaxy size dark matter haloes, since the
largest main haloes here lie in the right mass range $\sim
10^{12}\mathrm{M}_{\sun}/h$.

Although very interesting, the subhalo occupation number
distribution cannot be analyzed in detail due to poor statistics of
the data. The probability distribution $P(N_h|M_{\mbox{\s main}})$ is
probably Poissonian, as proposed by Kravtsov et al. (\cite{kratso2}).
In principle, this could be tested by dividing the main haloes into
different mass intervals and calculating the number distributions
for each mass interval, but there are too few haloes in the
simulations to obtain reliable distributions.

\begin{figure}
 \centering
 \resizebox{\hsize}{!}{\includegraphics*[angle=-90]{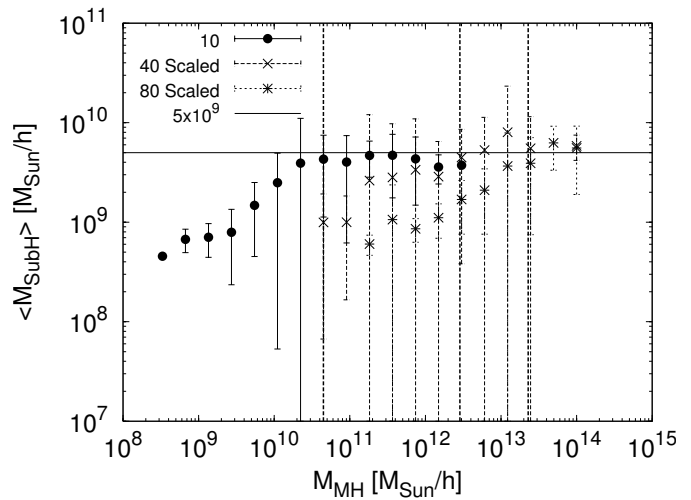}}
 \caption{Mean value of the subhalo
mass as a function of the main halo mass. The subhalo masses in the
B40 and B80 simulations are scaled with respect to the B10
simulation (see text for details). The mean subhalo  mass remains
constant in the reliable main halo mass region, over the whole mass
range covered by the simulations. The error bars show the standard
deviations of subhalo masses in the main halo mass intervals.}
\label{Kuva6_shmas}
\end{figure}

The subhalo mass function can be characterized also by the mean mass
of the subhaloes $\langle M_{SubH}\rangle$. It can be calculated
from the subhalo mass distribution $dn/dm$ and the total subhalo
mass fraction distribution (studied in the next section). The
$\langle M_{SubH}\rangle$ distribution for different main halo
masses depends obviously on the resolution, since the minimum
subhalo mass is related to the minimum mass of the main halo and
therefore it is determined by the particle mass used in the
simulation. Hence, the distribution of $dn/dm$ is cut at different
values, and without proper scaling, the minimum mass is larger for
larger volumes, where subhaloes are more massive (the N-body
particle mass is larger). Thus, we scaled the subhalo masses for the
40/$h$ and 80 Mpc/$h$ simulations by the particle mass ratios
between these and the 10 Mpc/$h$ simulation. The resulting
distributions of $\langle M_{SubH}\rangle$ are shown in
Fig.~\ref{Kuva6_shmas}. After this scaling $\langle M_{SubH}\rangle
$ is practically constant in the reliable main halo mass region; it
is an example of the self-similarity of different scales. This also
confirms earlier results by Moore et al. (\cite{moore}) and De
Lucia et al. (\cite{lucia}) that the mean subhalo mass is
independent of the main halo mass.

\subsection{Subhalo mass fraction}
In this section we analyze the mass fractions of subhaloes with
respect to their main haloes in detail. The motivation for that
comes from gravitational lensing studies. In general, strong
gravitational lensing (multiple quasar images and giant arc systems)
provides an unique way to study the dark matter content of galaxies
and galaxy clusters. Evidently, the dark matter substructure,
together with other characteristics of the deflecting lens, affects
the lensing cross section and thus the efficiency of lensing.

Numerical simulations play a crucial role for drawing quantitative
conclusions from lensing observations; the subhalo content of dark
matter haloes is especially important in this respect. Analytical
models do not properly take into account asymmetries in the lensing
mass distributions and systematically underestimate lensing
cross-sections (Torri et al. \cite{torri}, Meneghetti et al.
\cite{mene}). Due to the variations in the intrinsic properties of
lenses and to projection effects, variability of the model results
is large, and it is important to have large samples of simulations
to obtain reliable results. This is possible with a proper
combination of numerical simulations and analytical approximations,
which are used to overcome computational limitations.

\begin{figure}
 \centering
 \resizebox{\hsize}{!}{\includegraphics*[angle=-90]{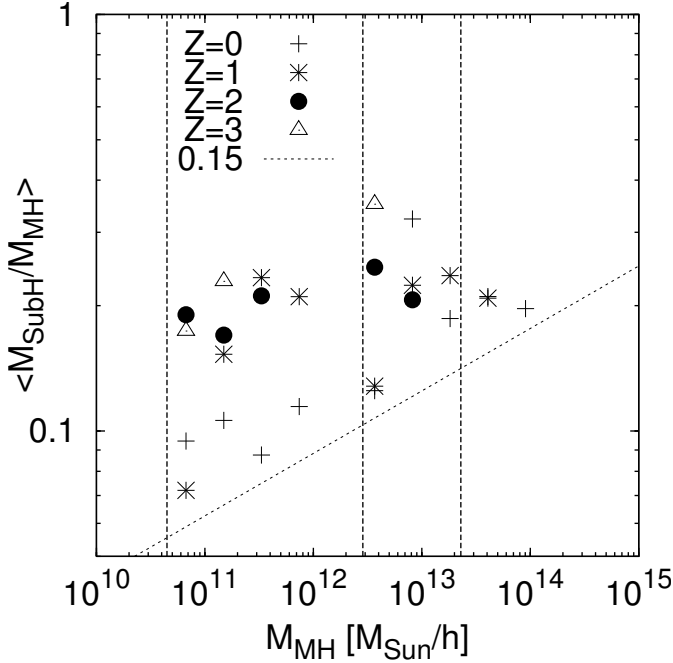}}
 \caption{Subhalo mass fractions at four different redshifts as functions
  of the main halo mass.} \label{mfs}
\end{figure}

First we study the dependence of the subhalo mass fraction on the
mass of the main halo. To illustrate how this mass fraction changes
in time, Fig.~\ref{mfs} shows the combined data (mean values of mass
fractions) for the three different simulations at four different
redshifts from $z=0$ to $z=3$. The vertical lines show the reliable
main halo mass regions for different simulations, and the resolution
limit $N_r=10^4$ is the same as before. Only main haloes in the
reliable regions are chosen for the analysis, and the number of main
haloes in a mass bin should be at least 4. This reduces statistical
fluctuations and makes general trends more evident. These general
trends become obvious at the redshifts 0 and 3. We illustrate this
trend in the figure; its functional dependence is given by $\langle
\Sigma M_{\mbox{\s Subh}}/M_{\mbox{\s MH}}\rangle\propto M_{\mbox{\s MH}}^{0.15}$.

The subhalo mass fraction at $z=0$ varies between 0.08 and 0.33, but
there is a general trend that this mass fraction is larger for more
massive main haloes than for small haloes. The trend is rather weak
for $z=2$ and also at $z=0$ and $z=1$ there are large statistical
fluctuations for some mass intervals. Despite these deviations, it
is an interesting evolutionary effect that we plan to study later.
Also, the subhalo mass fraction is generally larger ($\sim$ by a
factor of 2) at earlier redshifts ($z=3$ and $z=2$) than at the
present time, for the same main halo mass. Despite the scatter at
different mass bins the trend is systematic.

It is interesting to compare these results with mass fractions shown
in Gao et al. (\cite{gao}) and in van den Bosch et al.
(\cite{vanBosch05}). In Gao et al. (\cite{gao}) paper (their figure
7) a similar mass dependence is shown, but their mass fractions are
smaller than found in our study. Their mass fractions range between
0.06 and 0.09 for the main halo masses between
$3\times10^{13}$--$10^{15} \mathrm{M}_{\sun}/h$. A possible
explanation is that their subhalo masses are larger than in our
study and their main haloes are also more massive. Therefore, the
total mass fraction is smaller in their analysis. Van den Bosch et
al. (\cite{vanBosch05}) (their figure 8) also show how the subhalo mass
fraction varies as a function of redshift. According to their study,
there is a significant difference in this mass fraction between the
early and late epochs. At $z=3$ their mass fraction varies between
0.07 and 0.24, for the main halo masses from
$10^{11}\mathrm{M}_{\sun}/h$ to $10^{15}\mathrm{M}_{\sun}/h$. At
$z=0$ their mass fractions range from 0.02 to 0.08 in the same mass
interval. We do not find such a clear change in our simulations and
the difference of mass fractions for a certain main halo mass
interval is the same or even smaller for early redshifts, a results
that is actually opposite to that of van den Bosch et al.
(\cite{vanBosch05}). They used a semi-analytical model to compute the
masses of haloes and subhaloes. This fact may explain the
difference, at least partially, emphasizing the importance of the
comparison between different methods and different halo selection
algorithms.

\begin{figure}
 \centering
 \resizebox{\hsize}{!}{\includegraphics*{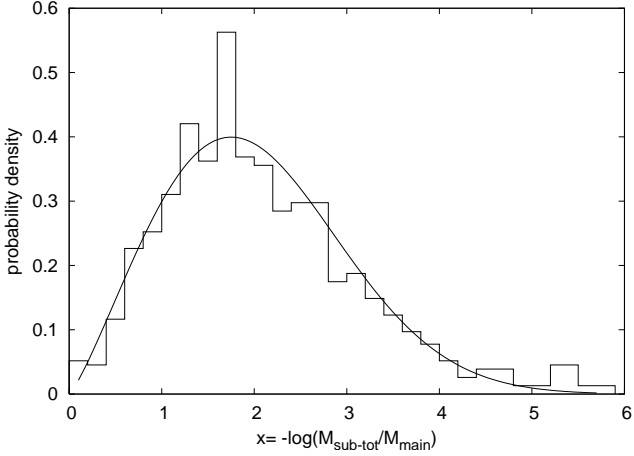}}
 \caption{Subhalo
mass fraction distribution for all simulations at the redshift
$z=0$; thick line shows the Weibull density fit.} \label{mf.z0}
\end{figure}

\begin{figure}
 \centering
 \resizebox{\hsize}{!}{\includegraphics*{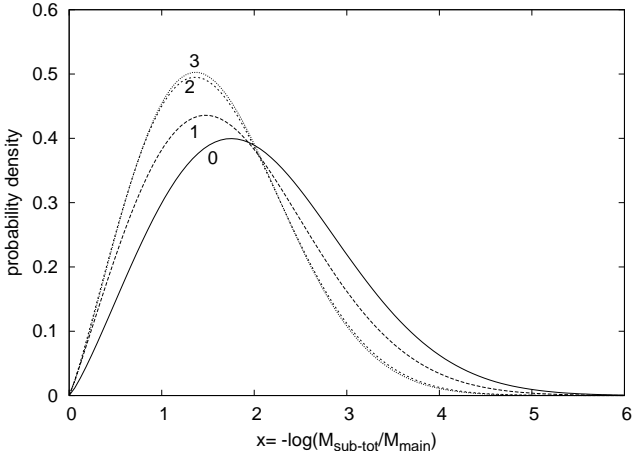}}
 \caption{Weibull-approximated  subhalo
mass fraction distributions for all simulations at the redshifts
$1-3$ (the curves are labeled by redshift values).} \label{mf.z1-5}
\end{figure}

We can carry out a more detailed analysis of the mass fraction
distributions, if we clump together all main haloes, ignoring their
mass. These distributions are shown in Figs.~\ref{mf.z0},
\ref{mf.z1-5}. We find that the shape of the distribution of the logarithm
of the mass ratio ($\log(M_{\mbox{\s Sub}}/M_{\mbox{\s MH}}))$
can be approximated by a Weibull distribution
(see, e.g., Evans, Hastings \& Peacock (\cite{distro})):
\[
f(x)=\frac{\gamma}{a}\left(\frac{x}{a}\right)^{\gamma-1}
  \exp\left(-(x/a)^\gamma\right).
\]


In the present case, the distribution describes well the 
overall shape of the observed mass ratio distributions, 
especially the small-ratio.

The $\chi^2$ fits give for the scale parameter $a$ values ranging
from 1.8--2.30, and for the shape parameter $\gamma$ a range of 2.08--2.21,
showing that the distributions at different redshifts are similar.
The fits are decent, with $\chi^2/ndf$ ($ndf$, as usual, is the
number of degrees of freedom) ranging from 1.4 to 2.1. The intrinsic scatter
of the histogram values was taken to be Poissonian, as customary.

Although the distribution parameters for different redshifts are close
to each other,
Fig.~\ref{mf.z1-5} shows that the mass ratio distribution evolves with
time. At earlier times (larger redshifts) the mass ratios were higher in 
the mean, and the small-ratio wing was not so heavy as at the present.
This is concordance with the picture of tidal disruption of subhaloes --
as the main halo evolves, subhaloes gradually lose their mass. 
We also see that this disruption is absent before $z=2$ (the distributions for
$z=3$ and $z=2$ practically coincide). 

These  distributions are useful for lensing studies.
It is known that lensing measurements are particularly sensitive to
the surface mass density distribution. Thus, to model lensing,
subhalo masses and their spatial locations in main haloes have to be
transformed to the projected surface density of the subhaloes, either
analytically or using numerical simulations.

The distributions shown in Figs.~\ref{mf.z0},\ref{mf.z1-5} 
provide a basis for
calculating the total mass fraction of subhaloes. These
distributions  can be used to find the radial number density
distributions of subhaloes, necessary  for generating the surface
mass densities for lensing studies.

\subsection{Halo environment}

The surface density of subhaloes depends, of course, on their
spatial density inside the halo. Also, subhaloes around the main halo
can contribute to this density, so it is useful to know the spatial
distribution of subhaloes at different distances.

\begin{figure}
 \centering
 \resizebox{\hsize}{!}{\includegraphics*[angle=-90]{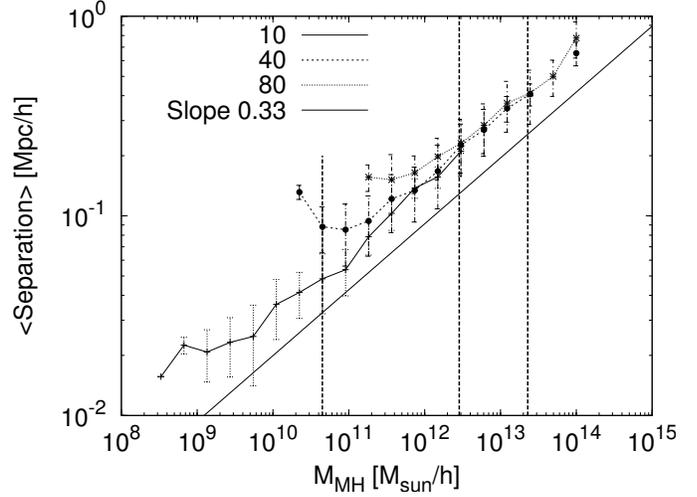}}
 \caption{The subhalo distance from its main halo center vs the main halo mass.}
\label{Kuva7_separation}
\end{figure}

Fig.~\ref{Kuva7_separation} shows the dependence of the mean
distance between the center of the main halo and its subhaloes on
the main halo mass. This figure can be easily explained, if we
assume that the mass of a halo $M_{\mbox{\s vir}}\propto
r_{\mbox{\s vir}}^3$. Since, for a universal subhalo distribution, the
ratio of the mean subhalo distance from the center  to the virial
radius of the main halo $\langle r\rangle/r_{\mbox{\s vir}}$ should be
constant, then $\langle r\rangle \propto r_{\mbox{\s vir}}$, and
therefore,  $\langle r\rangle \propto M_{\mbox{\s MH}}^{1/3}$, as shown
in Fig.~\ref{Kuva7_separation}.

\begin{figure}
 \centering
 \resizebox{\hsize}{!}{\includegraphics*[angle=-90]{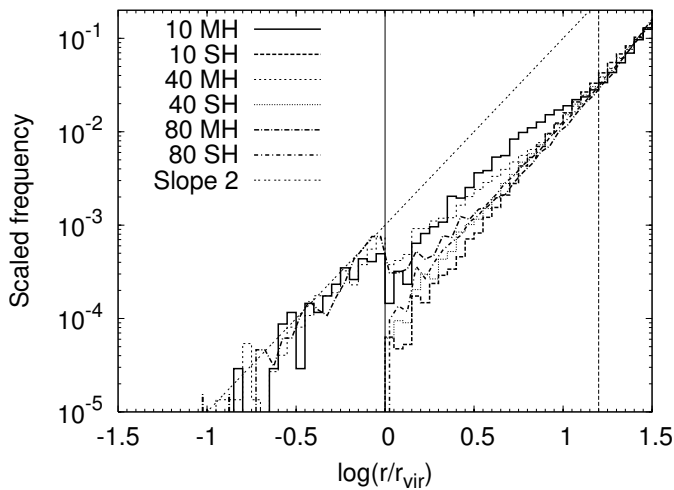}}
 \caption{The scaled number
distribution of haloes surrounding main haloes and single haloes in
all simulations at the redshift $z=0$. Haloes inside
$\log(r/r_{\mbox{\s vir}})\leq 1$ are subhaloes (solid line) and the
'sphere of influence' reaches up to $(r/r_{\mbox{\s vir}})=16$ (shown
as a dashed line).} \label{R_distribution2}
\end{figure}

\begin{figure}
 \centering
 \resizebox{\hsize}{!}{\includegraphics*[angle=-90]{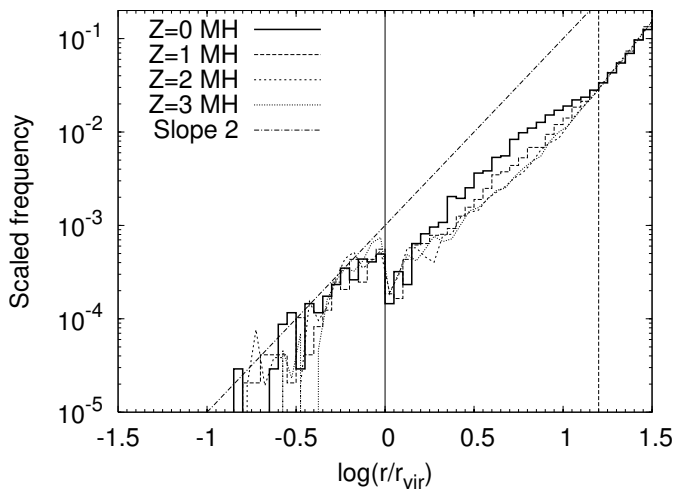}}
 \caption{The  number
distribution of haloes surrounding main haloes in the B10 simulation
at four different redshifts.} \label{R_distribution_z_Main}
\end{figure}

In order to study the environments of main haloes, we calculate the
number of surrounding haloes within a fixed separation interval as
we recede from the center of the halo. We scale separation distance
by the virial radius of the parent halo. By definition, all haloes
with $(r/r_{\mbox{\s vir}}) \leq 1$ are subhaloes. The number of
surrounding haloes decreases rapidly as we move out beyond this
limit. This is well seen in Fig.~\ref{R_distribution2}, that shows
the halo neighbor distribution for the redshift $z=0$; the data
about all main haloes in the three simulations are shown. The dashed
line of slope 2 shows the case of an uniform spatial distribution of
the haloes. In all cases this slope fits well in the subhalo region
and again, beyond the 'sphere of influence' that reaches out to the
distance $\log(r/r_{\mbox{\s vir}})\sim 1.2$. Our term 'sphere of
influence' describes the region beyond which the mean spatial
distribution of haloes does not evolve from $z=3$ to the present
epoch. This can be seen in Fig.~\ref{R_distribution_z_Main}. Inside
the sphere of influence the abundance of neighboring haloes is not
uniform. The difference is more evident in the case of the
simulation of the highest resolution, B10.
Fig.~\ref{R_distribution2} shows also the spatial distribution of
single haloes without subhaloes -- all haloes surrounding single
haloes are uniformly distributed. This means that single haloes
populate more poor environments than the haloes which contain
subhaloes. This can be explained to be a consequence of the fact
that single haloes are small on an average and they lie in less
dense regions.

To see how the halo environment varies as a function of redshift we
carried out the same analysis as before, but only for the B10
simulation (Fig.~\ref{R_distribution_z_Main}). We see that the
spatial distribution of subhaloes is almost the same for all
redshifts, although there is a large scatter. Studies with higher
resolution may reveal information about possible evolutionary
effects. However, beyond the subhalo region there is notable
difference between the spatial halo distributions of different
redshifts. At later redshifts the halo population is larger than
at early epochs. The Figure shows a rapid evolution of the halo
environment between $z=1$ and $z=0$, and a slower evolution at
earlier redshifts.

\begin{figure}
 \centering
 \resizebox{\hsize}{!}{\includegraphics*[angle=-90]{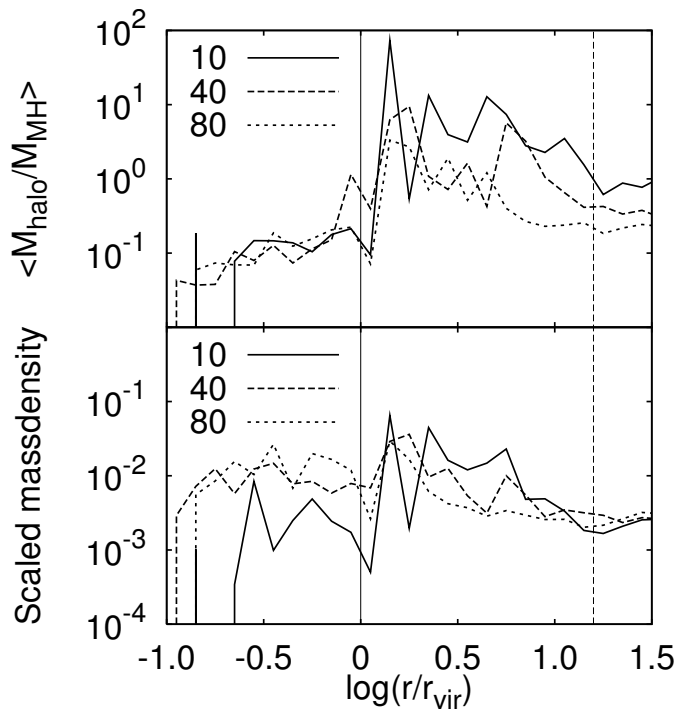}}
 \caption{
Mean value of the mass fraction for haloes around main haloes as a
function of separation for the B10, B40, and B80 simulations (upper
panel). The scaled mass density in the same mass bins is shown in
the lower panel.} \label{10_surroundings_mass}
\end{figure}

As a last issue, we will study how the mean values of the halo mass
fraction, calculated using all haloes surrounding the main haloes,
depends on the separation. This calculation was done for the
redshift $z=0$ and for the three simulations
(Fig.~\ref{10_surroundings_mass}, upper panel). In the lower panel
we show the scaled mass density at fixed $r/r_{\mbox{\s vir}}$ bins.
The scatter in both panels is much larger than it is in the number
distribution (Fig.~\ref{R_distribution_z_Main}), but the main trends
are still clear. As found by other authors (Reed et al.\cite{reed}
and references therein), subhaloes near the centers of their hosts
tend to have lower masses (smaller mass fractions) than subhaloes at
larger radii. The mean value of the mass fraction changes from 0.03
to 0.2 between $\log(r/r_{\mbox{\s vir}})=-1$ and 0. This is probably
due to substantial tidal stripping of subhaloes in the inner regions
of main haloes (De Lucia et al. \cite{lucia}, Nagai and Kratsov
\cite{nag:Kra}). Outside the subhalo region the mass fraction
increases rapidly (with large scatter), but after the peak value at
$\log(r/r_{\mbox{\s vir}}) \approx 0.25$ the mass fraction starts to
decrease and reaches a constant background level after
$(r/r_{\mbox{\s vir}})=16$ as in previous plots. The region inside
$r_{\mbox{\s vir}}$ is dominated by small main haloes that have only a
few subhaloes. Why the mass fraction rises rapidly beyond
$r_{\mbox{\s vir}}$? Since we calculate the mean value of the mass
fraction, we always have a few haloes that are more massive than the
mean mass of the main halo outside the virial radius, and therefore,
the mass fraction increases and there is a large scatter. If we look
at the density plot, we see that the mass density is only slightly higher
in this region than inside the virial radius, where it remains
constant. For the larger volumes ($r/r_{\mbox{\s vir}}>16$) the mean
mass reaches a level that is defined by the ratio of the mean mass
of the main haloes to the mean mass of all haloes inside the volume.
By numbers, this ratio is dominated by single haloes. We plan to
study the halo environments in more detail in later work.

In general, the study of the halo environment, the number of
subhaloes and their properties is a very interesting topic for
future studies. In addition, the halo 'sphere of influence' deserves
a deeper investigation that utilizes the information hidden in the
halo merging trees.

\section{Conclusion and Discussion}

We have analyzed the subhalo content in three cosmological
simulations calculated with different mass resolutions. The mass
functions for main haloes and subhaloes together with the mass
fraction distributions were found. We studied also the abundance of
haloes in the close vicinity of main haloes (up to 16 times
$r_{\mbox{\s vir}}$). Our conclusions are as follows:
\begin{enumerate}
    \item
By comparing theoretical and simulated mass functions of main haloes
and subhaloes we can set the limits for the minimum number of N-body
particles required to reliably select a halo. These limits are
$\sim100$ particles for subhaloes and $\sim10^4$ particles for main
haloes harboring subhaloes. These limits mark the masses at which
theoretical and simulated mass functions start to deviate. Within
the reliable region, practically all haloes harbor subhaloes, and
the existence of 'real' single haloes is questionable.
    \item
The functional form of the mass function of subhaloes agrees well
with earlier studies by Gao et al. (\cite{gao}) and Kravtsov et
al. (\cite{kratso2}). The subhalo mass function is the same in
different simulations, confirming the universality of the mass
function, since different simulations use different mass ranges and
mass resolutions.
    \item
The mass function slope is the same for main haloes and subhaloes,
but the slope is a function of redshift. The evolution of the slope
reflects the mass growth of haloes, being a consequence of the
complex accretion history of haloes.
    \item
The subhalo mass fraction depends on the main halo mass for
$M_{\mbox{\s MH}}$ between $10^{11}-10^{14}\mathrm{M}_{\sun}$, so that
more massive haloes have larger mass fractions. The subhalo mass
fraction at $z=0$ is between 0.08 and 0.2. Within the same main halo
mass range, the subhalo mass fraction is notably larger at earlier
epochs.
    \item
The distribution for the logarithm of mass fraction can be
approximated by a Weibull distribution. 
There is a systematic change in the distribution parameters 
as a function of redshift.
The dependence of the subhalo mass fraction on the
main halo mass depends on redshift, too, although not strongly.
    \item
The dependence of the number of subhaloes on the main halo mass can
be described by a simple relation
 $<N_h> \propto M_{\mbox{\s MH}}^{1.1}$.
    \item
The number density of haloes surrounding main haloes drops quickly
as we move beyond the virial radius of the halo. However, the slope
stays the same after that, up to distance $\sim3\times
r_{\mbox{\s vir}}$. The sphere of influence of a halo reaches out to
the distance of 16 times of its virial radius. Beyond this limit the
number density of haloes is uniform.
   \end{enumerate}

The knowledge of the fraction of the mass collapsed into bound
structures as a function of redshift is very important, as it links
simulations and observations. The information on the mass fraction
distributions and on differential mass functions, together with an
estimate for the scatter of values for different types of haloes,
can be used in modelling 'realistic' substructures around haloes.
This is of interest, for example, in different gravitational lensing
studies (Oguri \cite{oguri}, Mao et al. \cite{mao}).

In our simulations we found a large fraction of single haloes
($\sim$80\%), as opposed to main haloes that have subhaloes. This
fraction is smaller in higher resolution simulations and probably
the twofold character of such classification is ostensible. The
existence of 'real' single haloes is questionable. There is a large
scatter in the number of haloes within the same mass range, that
could be related to different halo environments. And, haloes with
substantially small number of subhaloes are found in lower density
regions.

The number of subhaloes varies significantly within a fixed main
halo mass interval, as was shown in this study. This cosmic
variation in the number of subhaloes can maybe explain the
sparseness of observed dwarf galaxies around some giant galaxies.
Especially, this could be true in low density regions, as in the
Local group, where dark energy might dominate the dynamical
evolution of haloes (Macci\`o et al. 2005 and Teerikorpi et al.
2005). This possible connection between the number of subhaloes and
matter density is certainly interesting.

It is surprising that the number of (sub)haloes drops as soon as the
main halo virial radius is reached. The drop is very clear and it
may reflect the dynamical effects of subhaloes, or it might be an
artefact of the halo identification algorithm. In the next study we
shall analyze in detail the environmental effects of haloes and shall
concentrate on the properties of individual haloes. Also, the
dependence of the abundance of subhaloes on their formation time is
an interesting subject for study.

The predicted change of the slope of the subhalo mass function as a
function of redshift could, in principle, be tested by compiling
observational mass functions of galaxy clusters and groups. However,
even the largest current surveys of galaxy clusters (for example,
the REFLEX cluster catalogue: B\"ohringer et al. (\cite{boh04})~ or
2df groups: Tago et al. (\cite{tago06})) are not deep enough; also,
the cosmic scatter in the propeties of individual clusters is large,
making the detection of evolution of the mass function difficult.

Finally, we demonstrated that information from the simulations,
using different mass resolutions, can be used efficiently to cover a
wide mass range of haloes, but one must be careful when the halo
content of haloes with less than $10^4$ particles is interpreted.
Simulations with different mass resolutions reveal important
information about the reliability of model haloes in different mass
ranges.

\begin{acknowledgements} This project was supported by the Finnish
Academy funding and by the Estonian Science Foundation grants
No.~4695 and 6104, and the Estonian Ministry for  Education and
Science grant TO 0060058S98. 
We also acknowledge support by the University of Valencia
through a visiting professorship for Enn Saar, and by the Spanish MCyT
project AYA2003-08739-C02-01 (including FEDER).
The cosmological simulations were run
at the CSC -- Scientific Computing center in Finland.
\end{acknowledgements}

\end{document}